\documentclass[aps,prl,twocolumn,showpacs]{revtex4}
\usepackage{epsfig}
\usepackage{graphicx}
\usepackage{amsmath}
\usepackage{amsfonts}
\usepackage{amssymb}
\usepackage{bm}
\usepackage{lineno}

\begin{document}
\title{Spin-Peierls instability in the three-leg Heisenberg ladder}
\author{Mohamed Azzouz}
\affiliation{Department of Physics, Laurentian University, Ramsey Lake Road,
Sudbury, ON, P3E 2C6, Canada }
\author{Khalada Shahin}
\affiliation{Department of Physics, Laurentian University, Ramsey Lake Road,
Sudbury, ON, P3E 2C6, Canada }
\author{Gennady Y. Chitov}
\affiliation{Department of Physics, Laurentian University, Ramsey Lake Road,
Sudbury, ON, P3E 2C6, Canada }

\date{\today}

\begin{abstract}
Because the three-leg ladder behaves
like a renormalized single Heisenberg chain
we argue that a spin-Peierls instability must occur in
this system when it is coupled to three-dimensional phonons.
Using the bond-mean-field theory, we show that this is indeed the case.
The dimerized state below the spin-Peierls transition
temperature forms into the columnar dimerized phase not the staggered
one. This contrasts with the argument based on antiferromagnetism.
A physical argument
based rather on spin bonding into singlets
explains why the columnar configuration is favored.
No quantum criticality (gaplessness) can occur
in the columnar arrangement of the dimerized chains.
\end{abstract}


\maketitle

Interest in the Heisenberg ladders stems from one
of their most intriguing aspects: namely, the critical properties are
dependent on the number of legs, and also from the various
(or at least potential) exotic quantum states
that can be realized in them (for a review
see Refs. \cite{Dagotto,Giamarchi04}.
The spin excitations in the
$n$-leg ladders with an even number $n$ of legs
are gapped, while those with  an odd
$n$  have gapless excitation spectra \cite{Dagotto}.
The even-$n$-leg ladders
provide an example of spin liquids.
The latter, defined (loosely) as gapped phases
without long-ranged order (LRO),
are, in particular, believed to be relevant
to the physics of high-$T_c$
superconductivity \cite{Lee06}.

In this Letter we study the effect of phonons on the
three-leg ladder. Having an odd number of
legs, this sytem is gapless in the absence of coupling to
phonons. This was confirmed, e.g., by the Monte Carlo simulations
\cite{Johnson00,Frisch96}. In the limit of strong rung coupling the three-leg
ladder behaves as an effective (renormalized) Heisenberg chain
\cite{Dagotto}. A recent thorough theoretical study on this system
corroborated this point \cite{Kofi}. Experiments on the real three-leg ladder
compound $\mathrm{Sr}_2 \mathrm{Cu}_3 \mathrm{O}_5$ find its spin
susceptibility similar to that of a (gapless) Heisenberg chain \cite{Azuma}.
It is therefore natural
to predict that a spin-Peierls (SP) instability can take place
in the three-leg ladder when it is coupled to phonons.
When a spin-$1/2$ Heisenberg chain is
coupled to phonons,
lattice distortions and a spin gap occur simultaneously below a
transition temperature $T_{\rm{SP}}$ known as the SP temperature
\cite{Giamarchi04}.
The gap is due to the distortion-induced
dimerization. The SP systems have been extensively studied both
experimentally and theoretically
\cite{Pytte,Cross,Orignac,azzouz1996}.
The SP instability has been observed in  ${\rm TTF-CuBDT}$,
${\rm TTF-AuBDT}$~\cite{Bray,Jacobs},
in other organic compounds \cite{Orignac},
and in the inorganic compound  CuGeO$_3$ \cite{Hase}.

\begin{figure}
\epsfig{file=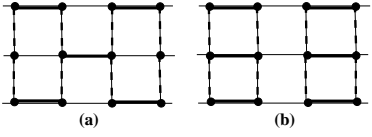,width=0.275\textwidth,angle=0}
\caption{
The bold/thin solid lines represent the stronger/weaker chain
coupling $J_0(1 \pm \delta)$, respectively. The dashed lines correspond to  the
rung coupling $J_\bot$. The dimerization pattern is staggered in (a)
and columnar in (b).}
\label{3LDim_Lad}
\end{figure}

We use the
Jordan-Wigner (JW) transformation \cite{Azzouz93} and bond mean-field
theory (BMFT) \cite{azzouz1994,azzouz1996,azzouz1997} to show that indeed
a SP transition takes place in the three-leg ladder when
it is coupled to three-dimensional phonons.
To the best of our knowledge, there are no other theoretical studies of this
problem, and real three-leg ladders with
an SP instability do not exisit yet.
There are, however, studies of the \textit{intrinsically dimerized} ladders.
According to Refs.~\cite{Delgado96,Almeida07}, for the antiferromagnetic
three-leg ladder with dimerization preset
in the staggered pattern (cf. Fig.\
\ref{3LDim_Lad}a), there exist a critical line in the
dimerization-rung-coupling plane ($\delta, J_\bot$),
where the system is gapless. Dimerized ladders provide a counterintuitive
example of ``restored quantum criticality", when a system (ladder) built from
gapped blocks (dimerized chains)
can be gapless, contrary to naive expectations. See
Refs. \cite{Delgado96,Almeida07} and Refs. therein for more detail.
Our analysis shows that this
interesting phenomenon does not occur in the coupled spin-phonon ladder: when
the ladder is allowed to choose
the dimerization pattern from the minimum energy
condition, it orders into the columnar phase (cf. Fig.\ \ref{3LDim_Lad}b). In
the latter case, the ladder of dimerized chains
is always gapped \cite{GBM}. So,
the SP transition in a three-leg ladder qualitatively resembles that in a
single chain.

The spin Hamiltonian for the three-leg ladder with antiferromagnetic couplings
and coupled to phonons  is
\begin{equation}
\label{hamil}
 H_{3L}= \sum_{i=1}^N \bigg[\sum_{j=1}^3 J_{i,i+1}(j) {\bf S}_{i,j}\cdot {\bf
 S}_{i+1,j}
+J_\bot \sum_{j=1}^2{\bf S}_{i,j}\cdot{\bf S}_{i,j+1}\bigg],
%
\end{equation}
where $i$ is the site label along the chains (i.e., rungs),
$j=1,2,3$ labels the legs (chains), and $N$ is the number of sites
in a single chain; the total number of spins is
$N_t=3N$. $J_\perp$ is the coupling along the rungs and
$J_{i,i+1}(j)$ is the longitudinal position-dependent coupling
because  the chains of the ladder are linearly coupled to
the phonon field $u_{i,i+1}(j)$
\cite{Giamarchi04}. Restricting our
analysis to the static alternating
lattice deformations $u$ along
the chains, we take
\begin{equation}
\label{Jeff}
J_{i,i+1}(j)=
J_0(1+\gamma \langle u_{i,i+1}(j) \rangle),~~\langle u_{i,i+1}(j) \rangle
\propto (-1)^i u
\end{equation}
In this (adiabatic) approximation the phonon
Hamiltonian is given by its static
deformation part
\begin{equation}
\label{eq:phonon1}
H_{\mathrm{ph}}= \frac12 N_t Ku^2 \equiv
\frac12 \frac{N_t J_0 \delta^2}{\lambda},
\end{equation}
where the dimensionless dimerization parameter
$\delta \equiv\gamma u$ and  the spin-phonon coupling $\lambda \equiv {J_0
\gamma^2 / K} $. The alternated frozen intrachain displacements (\ref{Jeff})
result in dimerization of each of the three chains. The dimerization of the
whole ladder can be in the staggered or columnar patterns, as shown in Fig.\
\ref{3LDim_Lad}.
%
%

The total Hamiltonian assumes then the form
$H=H_{\mathrm{3L}}+H_{\mathrm{ph}}$,
and the effective intrachain spin
coupling dependent on the dimerization pattern is
\begin{eqnarray}
\label{Jst}
 J_{i,i+1}(j) &=& J_0 [1+(-1)^{i+j} \delta ]~~\mathrm{staggered~} \\
\label{Jcol}
  J_{i,i+1}(j) &=& J_0 [1+(-1)^{i} \delta]~~~~~\mathrm{columnar}~.
\end{eqnarray}

The BMFT we apply to the JW fermions
consists of two key approximations. First, in
dealing with the phases resulting from the
JW transformation, the phase
differences due to hopping of the JW fermions around any given elementary
plaquette is set to be equal to $\pi$. Second, the quartic
JW fermionic terms
$c_{i,j}^\dag c_{i,j}c_{i+1,j}^\dag c_{i+1,j}$
resulting from the Ising interactions are decoupled using the bond
parameters \cite{Azzouz93,azzouz1997,Kofi,azzouz1994,azzouz1996}.
The latter are defined as  $Q_+=\langle
c_{2i,j}c_{2i+1,j}^\dag\rangle$,  $Q_-=\langle
c_{2i+1,j}c_{2i+2,j}^\dag\rangle$ for $j=1,3$; and  $Q'_+=\langle
c_{2i,2}c_{2i+1,2}^\dag\rangle$, $Q'_-=\langle
c_{2i+1,2}c_{2i+2,2}^\dag\rangle$ for $j=2$. For chains 1 and 3, the same
bond parameters are used because the ladder is symmetric
under exchanging chain labels 1 and 3 \cite{Kofi}.
In the direction along the rungs, only one bond parameter
is sufficient, $P=\langle c_{2i,j}c_{2i,j+1}^\dag\rangle$. For the reasons to
be explained below, we will concentrate on the columnar dimerization pattern
(\ref{Jcol}). Fourier transforming along the chains direction, keeping
the chains labels because of the open boundary conditions along the rungs,
and using the Nambu formalism, the BMFT yields
the single-particle effective Hamiltonian
\begin{equation}
\label{H}
  H^{(\rm co)} = \sum_k\Psi_k^\dag\mathcal{ H}^{(\rm co)} \Psi_k+C_t,
\end{equation}
where the Hamiltonian density $\mathcal{H}^{(\rm co)}$ is  a $ 6\times 6$
matrix given by
\begin{eqnarray}
\label{2ndmatrix}
 \mathcal{H}^{(\rm co)}
 = \left(\begin{array}{cccccc}
 0&{\mathcal{A}}&0&C&0&0\\
 {\mathcal{A}^{*}}&0&C&0&0&0\\
 0& C&0&{\mathcal{A}^\prime} &0&C\\
 C&0&{\mathcal{A}^{\prime *}}&0&C&0\\
 0&0&0&C&0&\mathcal{A}\\
 0&0&C&0&{\mathcal{A}^*}&0\\
\end{array}\right),
\end{eqnarray}
and the Nambu spinor ${\Psi_k^\dag}=
 \left(\begin{array}{cccccc} c_{1k}^{A\dag}&c_{1k}^{B\dag}& c_{2k}^{A\dag}&
 c_{2k}^{B\dag}& c_{3k}^{A\dag}& c_{3k}^{B\dag}
 \end{array}\right).$
Here $c_{jk}^{\alpha}$ is the Fourier transform of $c_{ij}^{\alpha}$
along the chain $j$; i.e., with respect to the index
$i$ ($\alpha=A,\ \rm {\rm or}\ B$), and because of the
antiferromagnetic correlations the lattice is
subdivided into two sublattices $A$
and $B$. The parameters entering the effective
Hamiltonian are: $\mathcal{A}^\sharp
=(J_{1+}^{\sharp} e^{ik}-J_{1-}^{\sharp} e^{-ik})/2$, and
$C=J_{\bot1}/2$, where
\begin{eqnarray}
\label{AD}
 J_{1\pm}^\sharp  &=&J_0(1\pm\delta)(1+2Q_\pm^\sharp )~, ~
 J_{\bot 1}= J_\bot(1+2P),\cr
 C_t &=& NJ_+| Q_+|^2+ NJ_- |Q_-| ^2+ \frac{N}{2} J_- |Q'_-|^2 \cr
 &+& \frac{N}{2}J_+ |Q'_+|^2+2J_\bot N |P|^2+\frac{3NJ_0\delta^2}{2 \lambda }.
\end{eqnarray}
Here, $J_{1\pm}^\sharp=J_{1\pm}$ for $Q_\pm^\sharp=Q$ or $J_{1\pm}'$ for
$Q_\pm^\sharp=Q'$.
Diagonalizing $\mathcal{H}^{(\rm co)}$ yields six energy eigenvalues $ \pm
E^{(\rm co)}_j(k)$, $j=1,2,3$, where
\begin{equation}
\label{E1}
    E^{(\rm co)}_1(k) = \frac{1}{2}\sqrt {z},
\end{equation}
and
\begin{widetext}
\begin{equation}
\label{E23}
 E^{(\rm co)}_{n}(k)= \frac{1}{2^{\frac32} } \left[(-1)^n\sqrt
 {(z-z')^2+8J_\bot^2(z+z')+16J_{\bot 1}^2t}+z+z'+4J_{\bot 1}^2\right]
^\frac{1}{2}, ~~n=2,3.
\end{equation}
\end{widetext}
In Eqs.~(\ref{E1},\ref{E23})
\begin{eqnarray}
\label{eq:zt}
 z^\sharp &\equiv& (J_{1+}^\sharp)^2+ (J_{1-}^\sharp)^2
- 2 J_{1+}^\sharp J_{1-}^\sharp \cos(2k),\cr
 t &\equiv& \big(J_{1+}J'_{1-}+J_{1-}J'_{1+}\big)\cos
(2k)-\big(J_{1+}J'_{1+}+J_{1-}J'_{1-}\big).\nonumber
\end{eqnarray}
In the absence of dimerization  ($\delta=0$),  the energy eigenvalues
(\ref{E1},\ref{E23}) coincide with those found in Ref.~\cite{Kofi}.

The partition function of the single-particle Hamiltonian (\ref{H}) can be
calculated and leads to the following free energy per spin
\begin{equation}
\label{FrEn}
 F^{(\rm co)}=\frac{C_t}{3N}-\frac{1}{
 2 \beta N_t}\sum_{p=\pm}\sum_k\sum_{j=1}^3
\ln[1+e^{p\beta E^{(\rm co)}_j(k)}],
\end{equation}
where $\beta=1/ {k_BT}$. Finally, the mean-field equations are derived from
minimization of the free energy (\ref{FrEn}) with respect to the six mean-field
parameters $Q_\pm, Q'_\pm$, $P$ and $\delta$. These self-consistent (integral)
equations are solved numerically.

The mean-field equations  predict no magnetic long range order (LRO), even at
zero temperature. At the same time, they predict the simultaneous appearance at
some critical temperature of the structural LRO (lattice dimerization $\delta$)
and the spin gap, generated by $\delta \neq 0$. So, this is a SP
transition.

\begin{figure}
\epsfig{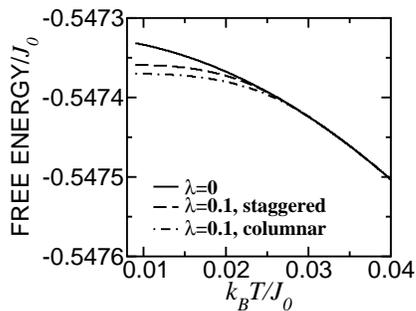}
\caption{The free energies of the three-leg ladder without phonons
($\lambda=0$, no dimerization), and in the presence of phonons ($\lambda=0.1$)
for the staggered and columnar configurations are
plotted as functions of ftemperature.}\label{F3}
\end{figure}

\begin{figure}
\epsfig{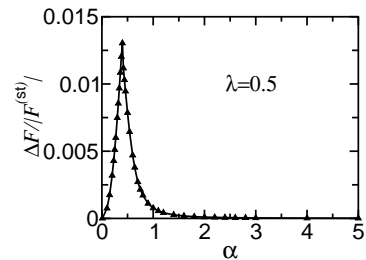}
\caption{The free energy difference between the staggered and columnar
configurations as a function of $\alpha\equiv J_{\bot}/J_0$ at $T=0.009J_0$,
$\lambda =0.5$.}\label{DeltaF}
\end{figure}

First, we compare the free energies of the two dimerization patterns. For the
staggered configuration with the effective coupling (\ref{Jst}), the
single-particle effective Hamiltonian, its spectrum, and the mean-field
equations are derived in the same manner as described above for the columnar
configuration. Fig.~\ref{F3} shows the free
energies as functions of temperature for both
dimerized configurations and without dimerization ($\lambda=0$).
The free energy of the dimerized columnar configuration is lower than those of
the staggered or non-dimerized configurations below a nonzero temperature,
which we identify as the SP temperature $T_{SP}$.
So, the columnar order constitutes the
thermodynamically stable state at low temperature. Above $T_{SP}$
the dimerization disappears, and all
three free energies become equal.
%
%
%
%
%
We also plot $\frac{\Delta F}{\mid F^{(\rm{st})}\mid}=
\frac{F^{(\rm{st})}-F^{(\rm{co})}}{\mid F^{(\rm{st})}\mid}$ as a function of
$\alpha\equiv J_{\bot}/J_0$ at practically zero temperature in Fig.~\ref{DeltaF}
for $\lambda=0.5$.
$\frac{\Delta F}{\mid F^{(\rm st)}\mid}$ increases as $\alpha$
increases passes through a maximum, then decreases rapidly in the strong
coupling regime $\alpha\gg 1$.
This means that the
staggered and columnar states become practically degenerate when $\alpha\gg1$.
%
%

To understand why the columnar configuration is the stable one, assume
$\alpha$ is large. In this limit, the three-leg ladder
behaves pretty well like a single Heisenberg chain
with an effective coupling, and the spin degrees of freedom
on two of the chains freeze into spin singlets
on the rungs \cite{Kofi}. It is then favorable for the
coupling along the chains to couple the singlets
into a plaquette (which gives rise to
a linear combination of transverse
and longitudinal spin singlets for the spins on four of the six sites)
like in
Fig.~\ref{3LDim_Lad}(b). The staggered configuration would be
favorable if the spins on the rungs were
ordered antiferromagnetically.
For intermediate and weak values of the rung coupling $\alpha$,
the spins are not frosen but are
nonetheless locked into random singlets on two of
the three chains \cite{Kofi}, and the argument based on spins singlets
still holds, while that based on antiferromagnetically ordered spins
on the rungs continues to be wrong.

\begin{figure}
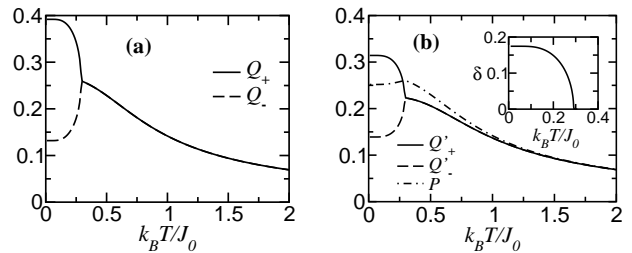

\includegraphics[height=3.25cm]{Fig4a.eps}
\hspace{.3cm}
\includegraphics[height=3.25cm]{Fig4b.eps}
\caption{The temperature dependence of the mean-field parameters $Q_\pm$,
$Q'_\pm$, $P$ and $\delta$ for  $\alpha=1$, $\lambda=1$.}\label{parameters}
\end{figure}

The mean-field parameters are plotted as functions of
temperature in Fig.~\ref{parameters} for $\lambda=1$ and $\alpha=1$.
At $T_{SP}$, a spin gap opens and a
structural phase transition occurs from the disordered phase
into the dimerized phase. This is clearly indicated by
the temperature dependence of $\delta$ in the inset of figure~\ref{parameters}
(b), which is very similar to that in an ordinary second-order
phase transition. The other (bond) parameters are not critical.
%
%
Below $T_{SP}$, $Q_+^\sharp \neq Q_-^\sharp$ due
to non-zero dimerization, while $Q_+^\sharp = Q_-^\sharp$ above $T_{SP}$.

\begin{figure}
\epsfig{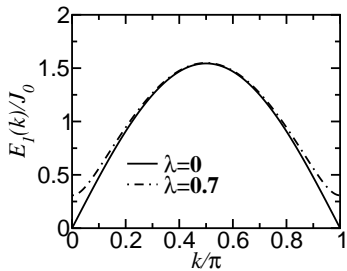}
 \caption{The spectra for the uniform ($\lambda=0$) and columnar dimerized
($\lambda=0.7$) ladders are shown. Here, $\alpha=1$.}\label{kofeandshahin}
\end{figure}

\begin{figure}[b]
\includegraphics[height=3.9cm]{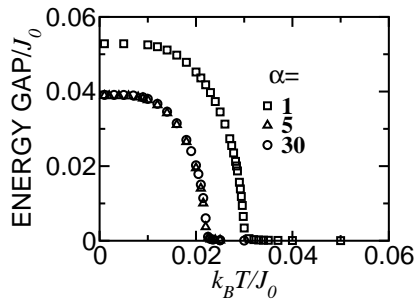}
\caption{ The temperature dependence of the spin
gap is shown for several values
of $\alpha$, and for
$\lambda=0.1$.}\label{gapgraph}
\end{figure}

The SP instability is accompanied by the opening
of an energy gap in the spin excitation
spectra (spin gap). The spin excitation energies consist of three bands
$E_j(k)$,  Eqs. (\ref{E1}), (\ref{E23}), with $E_2$ and $E_3$ gapped
even in the absence of dimerization with
a gap of the order of $J_{\bot}$. The gap in $E_1(k)$ is induced by the
dimerization $\delta$. We plot  $E_1(k)$ for the columnar dimerized
$(\lambda=0.7)$ and non-dimerized $(\lambda=0)$ ladder in
Fig.~\ref{kofeandshahin}. The spectrum of the non-dimerized three-leg ladder,
calculated in Ref.~\cite{Kofi} is gapless, i.e.,
$E_1(0/ \pi)=0$, as it should be
for an odd-leg ladder \cite{Dagotto}. (Note the symmetry of $ E_j(k)$ with
respect to the point $k=\frac{\pi}{2}$.)
%
%
%
%
The spin gap $E_g \equiv E_1(0)$ defined from
Eqs. (\ref{AD}) and (\ref{E1}) is
\begin{equation}
\label{SG}
E_g= J_0 \big[ \delta (1+Q_+ +Q_-)+ (Q_+ - Q_-) \big].
\end{equation}
So, to leading order our theory, as any mean-field one
\cite{Bulaev63} gives the
non-interacting fermion (equivalently, the dimerized single $XY$-chain) result
$E_g \propto \delta$, albeit renormalized by the bond parameters
$Q_{\pm}$. This is confirmed by direct numerical calculations of
$E_g(\delta)$. Note that for a dimerized chain $E_g
\propto \delta^{2/3}$ \cite{Cross}, corrected by the marginal logarithmic
prefactor \cite{BE81,Affleck89,Orignac}. The temperature dependence of the spin
gap is shown in Fig.~\ref{gapgraph}.
Both the numerical calculations and qualitative analyses of the mean-field
equations show that $\delta$, $E_g$, and
$T_{SP}$ increase monotonously with the phonon coupling constant $\lambda$.
However the dependence of those parameters on
$\alpha = J_{\bot}/J_0$ is trickier. The parameters $\delta,\ E_g$ and $T_{SP}$
initially decrease with growing $\alpha$, but then saturate as $\alpha \gtrsim
5$ as seen in Fig.~\ref{gapgraph}. The mechanism of such saturation was
already discussed \cite{Dagotto, Kofi}, and can be understood in
the regime $\alpha \gg 1$ as due to the fact that the ladder behaves as a
``renormalized" spin-$\frac12$ chain.

In this work, the spin-Peierls instability in the three-leg ladder coupled to
phonons is studied. We map the spin operators of the three-leg Heisenberg
ladder onto Jordan-Wigner fermions. The resulting interaction terms are
decoupled within the bond-mean-field theory. This theory yields a
spin-Peierls transition into the columnar dimerized phase.
Qualitatively, the three-leg dimerized ladder behaves like a single
spin-Peierls chain.  No gapless state can occur for such arrangement of the
dimerized chains. Finally, we conjecture that the results derived here
for the three-leg ladder should stay true for any $n$-leg
ladder with $n=5,\ 7,...$
%
%
%
%
\begin{acknowledgments}
We acknowledge financial support from the Natural Science and Engineering
Research Council of Canada (NSERC) and the Laurentian University Research Fund.
\end{acknowledgments}
%


\begin{thebibliography}{}
%
\bibitem{Dagotto} E. Dagotto and T. M. Rice Science \textbf{271}, 618 (1996).
%
\bibitem{Giamarchi04} T. Giamarchi,
\textit{Quantum Physics in One Dimension} (Oxford University Press, Oxford,
2004).
%
\bibitem{Lee06} P.A. Lee, N. Nagaosa, and X.-G. Wen,
Rev. Mod. Phys. \textbf{78}, 17 (2006).
%
\bibitem {Johnson00}
D.C. Johnston, M. Troyer, S. Miyahara, D. Lidsky, K. Ueda, M. Azuma, Z. Hiroi,
M. Takano, M. Isobe, Y. Ueda, M.A. Korotin, V.I. Anisimov, A.V. Mahajan, L.L.
Miller, cond-mat/0001147.
%
\bibitem {Frisch96}
B. Frischmuth,  B. Ammon, and  M. Troyer, Phys. Rev. B \textbf{54}, R3714
(1996).
%
\bibitem{Kofi} M. Azzouz and K. A. Asante, Phys. Rev. B \textbf{72}, 094433 (2005).
%
\bibitem{Azuma} M. Azuma, Z. Hiroi, M. Takano, K. Ishida and Y. Kitaoka,
Phys. Rev. Lett. \textbf{73}, 3463 (1994).
%
\bibitem{Pytte} E. Pytte, Phys. Rev. B \textbf{10} 4637 (1974).
%
\bibitem{Cross} M. C. Cross and D. S. Fisher, Phys. Rev. B \textbf{19}, 402 (1979).
%
\bibitem{Orignac} E. Orignac and R. Chitra,
Phys. Rev. B \textbf{70}, 214436 (2004).
%
\bibitem{azzouz1996} M. Azzouz and C. Bourbonnais,
Phys. Rev. B {\bf 53}, 5090 (1996) for the SP instability in higher dimensions.
%
\bibitem{Bray} J. W. Bray, H. R. Hart, Jr.,
L. V. Interrante, I. S. Jacobs, J. S. Kasper,
G. D. Watkins, S. H. Wee, and J. C. Bonner, Phys. Rev. Lett.  \textbf{35}, 744
(1975).
%
\bibitem{Jacobs}
I. S. Jacobs, J. W. Bray, H. R. Hart, Jr., L. V. Interrante, J. S. Kasper, G.
D. Watkins, D. E. Prober, and J. C. Bonner, Phys. Rev. B \textbf{14}, 3036
(1976).
%
\bibitem{Hase} M. Hase, I. Terasaki, and K. Uchinokura, Phys. Rev. Lett.
\textbf{70}, 3651 (1993).
%
\bibitem{Azzouz93} M. Azzouz, Phys. Rev. B \textbf{48}, 6136 (1993).
%
\bibitem{azzouz1994} M. Azzouz, L. Chen, and S. Moukouri,
Phys. Rev. B {\bf 50}, 6233 (1994).
%
\bibitem{azzouz1997}
M. Azzouz, B. Dumoulin, and A. Benyoussef, Phys. Rev. B {\bf 55}, R11957 (1997).
%
%
\bibitem{Delgado96} M.A. Martin-Delgado, R. Shankar, and G. Sierra,
Phys. Rev. Lett. \textbf{77}, 3443 (1996).
%
\bibitem{Almeida07} J. Almeida, M.A. Martin-Delgado, and G. Sierra,
arXiv:0707.4452 (2007).
%
\bibitem{GBM} The criticality of the various ordered dimerized ladders
will be addressed elsewhere: G.Y. Chitov, B. Ramakko, and M. Azzouz, \textit{in
preparation}.
%
\bibitem{Bulaev63} L.N. Bulaevskii, Zh. Eksp.
Teor. Fiz. \textbf{44}, 1008 (1963)
[Sov. Phys. JETP \textbf{17}, 684 (1963)].
%
\bibitem{BE81} J.L. Black and V.J. Emery,
Phys. Rev. B \textbf{23}, 429 (1981).
%
\bibitem{Affleck89}
I. Affleck, D. Gepner, H.J. Schulz, and T. Ziman, J. Phys. A {\bf 22}, 511
(1989).
%

%
\end{thebibliography}
\end{document}